\documentclass[12pt]{article}
\usepackage{amsfonts}
\usepackage{xcolor}
\usepackage{graphicx}
\usepackage{color,soul}
\usepackage{amsmath}
\usepackage{float}
\textwidth=6.5in \hoffset=-.75in \voffset=-1in \numberwithin{equation}{section}
\numberwithin{figure}{section}

\newcommand {\be}{\begin{equation}}
\newcommand {\ee}{\end{equation}}

\begin{document}

\begin{titlepage}
\vspace{1cm}
\begin{center}
{\Large \bf {Fubini-Study geometries in the higher-dimensional gravity}}\\
\end{center}
\vspace{2cm}
\begin{center}
A. M. Ghezelbash{ \footnote{ E-Mail: masoud.ghezelbash@usask.ca}}
\\
Department of Physics and Engineering Physics, \\ University of Saskatchewan, \\
Saskatoon, Saskatchewan S7N 5E2, Canada\\
\vspace{1cm}
\vspace{2cm}
%\today\\
\end{center}

\begin{abstract}
We construct approximate solutions to the Einstein-Maxwell theory with uplifting the four dimensional Fubini-Study Kahler manifold. We find the solutions can be expressed as the integrals of two special functions. The solutions are regular almost everywhere except a bolt structure on a single point in any dimensionality.
We also show that in the context of considered ansatzes for the metric function and the Maxwell field, the solutions are unique and can not be non-trivially extended to include the cosmological constant in any dimensions.

\end{abstract}
\end{titlepage}\onecolumn 
\bigskip 

\section{Introduction}

Finding the exact or approximate solutions to Einstein gravity as well its extensions, is essential to explore the nature of the gravitation and its properties in diverse phenomena in different dimensionality of the spacetime. In particular, 
embedding the lower-dimensional solutions of gravity in the higher-dimensional theories, which normally have more field degrees of freedom, enable us to generate new solutions. Analyzing the new solutions in turn, open the door to understand better the properties of the higher-dimensional theories , as well as their connections to the lower-dimensional gravity. 

By the way, new phenomena can emerge from the higher-dimensional theories that there are no counterparts in the lower-dimensional gravity.  Discovery of the black objects with non-trivial  horizon topologies in five dimensions  \cite{Myers1}--\cite{ El1} is just one example of some rich structures, which do not exist in four-dimensional  in gravity. A few  other interesting higher-dimensional solutions consist of solutions inspired by string theory \cite{EY5} and supergravity solutions \cite{other13}--\cite{hashi}. 

In fact, in \cite{hashi}, the authors constructed the convolution-like solutions in eleven-dimensional supergravity. Upon compactifying the convolution-like M2-brane solutions over a circle of transverse self-dual Taub-NUT geometry, they found the solutions for the the fully localized type IIA D2-branes intersecting D6-branes. An interesting feature of the solutions is that they preserve partial supersymmetries. Other interesting feature of the solutions is that they are valid everywhere; near and far from the core of D6-branes. The solutions are obtained completely in terms of convolution-like integrals, that is a result of taking a special ansatz for the solutions and separability of the equations of motion. Inspired by the convolution-like M2-brane solutions, in this article, we try to construct such convolution-like solutions in six and higher-dimensional Einstein-Maxwell theory by employing the special metric ansatzes (\ref{ds6}), (\ref{ds7}) and (\ref{mD}), that lead to convolution-like solutions.

Moreover in references  \cite{EY2}-\cite{ Be}, the dyonic, solitonic as well as instanton solutions to higher-dimensional extension of  gravity, including the Einstein-Maxwell theory, were constructed.  The construction involves uplifting one or more four-dimensional  base spaces into the higher-dimensional theory, by assuming simple ansatzes for the metric and other relevant fields in the theory. The success of construction largly relies on reducing the field equations to one or a few partial differential equations which are separable/solvable, and admit proper boundary conditions.  
 
We note that if the uplifted four-dimensional base space is a hyper-Kahler manifold, then the solutions of the  higher-dimensional theories of gravity (or their extensions) possess supersymmetry. In other words, the hyper-Kahlericity of the uplifted space (which is equivalent to a metric with self-dual curvature tensor)  guarantees, at least partially,  the existence of  full/partial supersymmetry in the higher-dimensional theory  \cite{GA}.
 
 In \cite{kumar}, the authors present few exact solutions in the higher dimensional Einstein-Maxwell theory, in which the metric functions are convoluted-like integrals of two special functions. The solutions are almost regular everywhere, and describe the wormhole handles in the higher dimensional Einstein-Maxwell theory. 
 
In this article, we consider the Fubini-Study metric in four dimensions, which is an exact solutions to the Einstein equations with a cosmological constant. The Fubini-Study metric has been used to construct the balanced metrics in the context of complex geometry \cite{hashimoto}, as well as classifying the non-isometric two-dimensional Kahler-Einstein sub-manifolds of a finite dimensional complex projective space \cite{Gianni}.  The  suggested deep connection between the quantum information theory can be established through the Fubini-Study metric \cite{Run, Carlo}. The space also has been used to define complexity in a quantum field theory \cite{Mario}.

In this article, we embed the Fubini-Study metric into the six and higher dimensional Einstein-Maxwell theory.  We find the solutions, in any dimensions  greater than or equal six, in the form of convolution-like integrals, and discuss the properties of the solutions. The convolution-like  solutions have been constructed in different theories of gravity, such as  M-theory \cite{MB}. 
 We note that  the minimal dimensionality of Einstein-Maxwell theory should be six, to have non-trivial convolution-like solutions. We also consider the Einstein-Maxwell theories with a cosmological constant in six and higher dimensions and  show explicitly there is no cosmological constant in any dimensions.

The article is organized as follows. In section \ref{sec:FS}, we review and discuss the physical properties of the Fubini-Study metric. In section \ref{sec:6D}, we present the  solutions to the six-dimensional Einstein-Maxwell theory. Following this, in section \ref{sec:7D}, we present the  solutions to the seven-dimensional Einstein-Maxwell theory. We use the results of the section  \ref{sec:7D}, to construct the more general solutions in the higher-dimensional theory in section \ref{sec:8D}. We also find that there are no solutions to the Einstein-Maxwell theory with cosmological constant if we consider a special anasatz for the metric, which is given by (\ref{mDLambda})  with a time-dependent metric function. Of course, we don't roll out finding other possible cosmological solutions with more sophisticated ansatzes. We wrap up the article by the  concluding remarks in section \ref{sec:con}.

\section{The Fubini-Study geometry }
\label{sec:FS}
The Fubini-Study metric in four dimensions  is given by
\be
ds^2_{FS}={\frac {{d{{r}}}^{2}+1/4\,{r}^{2} \left( d{{\psi}}+\cos \left( 
\theta \right) d{{\phi}} \right) ^{2}}{ f(r)^2}}+\frac{1}{4}\,{\frac {{r}^{2} \left( {d{{\theta}}}^{2}+
 \left( \sin \left( \theta \right)  \right) ^{2}{d{{\phi}}}^{2}
 \right) }{f(r)}}\label{metr},
\ee
where 
\be
f(r)= 1+\frac{1}{6}\,\Lambda\,{r}^{2}
.
\ee
The coordinate $\psi$ parameterizes the fibration of a circle over the sphere and $r \geq 0$.
 The cosmological constant $\Lambda$ in (\ref{metr}) can take any positive or negative values. The zero cosmological constant leads to a trivial fibration of a circle on an $S^2$. Though for the negative values of $\Lambda$, the metric function $f(r)$  has two zeros, however the Ricci scalar and the Kretschmann invariant are regular for all values of the radial coordinate $r$. In fact, the Ricci scalar is 
\be
R=4\Lambda\label{Ricci},
\ee
and the Kretschmann invariant $K$ is equal to
\be
K=\frac{16}{3}\Lambda^2\label{K}.
\ee
All the components of the Ricci tensor diverges at the zeros of the metric function $f(r)$, where  $\Lambda < 0$. 
The Fubini-Study space (\ref{metr}) is a special case ($N=2$) of the class of $2N$-dimensional Kahler manifolds over complex projective space ${\bold {CP}}^N$, with the affine coordinates $(z_1,\cdots,z_N)$. The metric tensor for the $2N$-dimensional space is given by
\be
g_{\mu{\bar \nu}}=\frac{{\delta _{\mu \bar \nu}}}{\sqrt {G(z_1,\cdots,z_N)} }-\frac{\bar z_\mu z_\nu}{G(z_1,\cdots,z_N)},\label{gmm}
\ee
where
\be
G(z_1,\cdots,z_N)=(1+\sum _{i=1}^N |z_i|^2)^2.
\ee
Using (\ref{gmm}), the line element for the 
 $2N$-dimensional Fubini-Study space can be written as
\be 
ds^2_{2N}=g_{\mu\bar \nu}dz^\mu d\bar z^\nu=\frac{dz_\mu d \bar z^\mu}{\sqrt{G}}-\frac{z_\mu \bar z_\nu dz^\nu d \bar z^\mu}{G}\label{FS2N},
\ee
The line element (\ref{FS2N}) also could be derived directly from the Fubini-Study Kahler potential 
${\cal K}$  
\be
e^{2\cal K}=G,
\ee
by
\be
ds^2_{2N}=\frac{\partial ^2 {\cal K}}{\partial_{z^\mu}\partial_{\bar z^\nu}}dz^\mu d\bar z^\nu.
\ee
We note that for $N=1$, the metric (\ref{FS2N}), reduces simply to the metric of a 2-sphere $S^2$. 
We also note that for $N = 2$, the curvature two-form for the Fubini-Study space is not self-dual, however the Weyl two-form is self-dual. More precisely the curvature two-form 
$R_{ab} = d\omega_{ab} + \omega_{ac}\wedge  \omega_b^c$ is not equal to its Hodge dual $\star R_{ab}$, where $\omega_{ab}$ is the Levi-Civita spin connection, and $a$ and $b$ are vierbein indices of the vierbeins $e^a$. On the other hand, the Weyl two-form $W_{ab} = \frac{1}{2}W_{abcd} e^c \wedge e^d$ is equal to its Hodge dual $\star W_{ab}$, where 
$W_{abcd}$ is the Weyl tensor.
 %%%%%%%%%%%%%%%%%%%%%%%%%%%%%%%%%%%%%%%%%%%%%%%%%%%%%%%%%%%%%%%%%%%%%%%%%%%%%%%%%%%%%%%%%%%%%%%%%%
\section{Embedding the four-dimensional Fubini-Study space in the six dimensional Einstein-Maxwell theory}
\label{sec:6D}

We embed the 4-dimensional Fubini-Study space (\ref{metr}) in the six-dimensional spacetime, which is  given by
\begin{equation}
ds^{2}=-\frac{dt^{2}}{H(r,x)^{2}}+H(r,x)^{2/3}(dx^2+ds_{FS}^2),
\label{ds6}
\end{equation}
where the coordinate $x\geq 0$.  We also consider the components of the electromagnetic field tensor, as
\begin{eqnarray}
F_{tr}&=& - \frac{\gamma}{H(r,x)^2}\frac{\partial H(r,x)}{\partial r} ,\label{F1}   \\  F_{tx}&=&- \frac{\gamma}{H(r,x)^2}\frac{\partial H(r,x)}{\partial x} 
\label{gauge},
\end{eqnarray}
where $\gamma$ is a constant. 

\begin{figure}[H]
\centering
\includegraphics[width=0.5\textwidth]{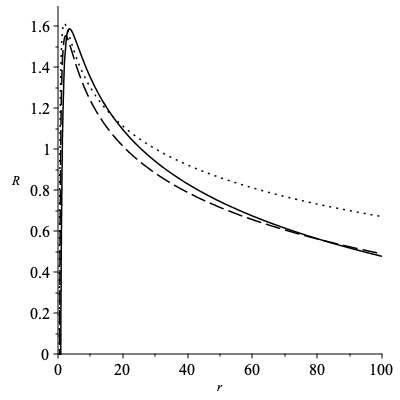}
\caption{The radial function $R(r)$ for different values of the cosmological constant $\Lambda=2$ (solid), $\Lambda=3$ (dash) and $\Lambda=4$ (dot), where we set $r_1=\frac{2}{\sqrt{6}}$, $r_2=0$, and $p=1$.}
\label{Fig1}
\end{figure}

The gravitational field equations in geometrized units, are given by 
\be
G_{\mu\nu}=T_{\mu\nu},\label{eq1}
\ee
where $G_{\mu\nu}$ is the Einstein tensor and $T_{\mu\nu}$ is the energy-momentum tensor for the electromagnetic field, which is given by
\be
T_{\mu\nu}=  F_{\mu}^{\lambda}F_{\nu\lambda}-\frac{1}{4}F^2g_{\mu\nu}.\label{T6}
\ee
Moreover, the electromagnetic field equations are given by
\be
F^{\mu\nu}_{;\mu}=0,\label{eq2}
\ee
A straightforward but lengthy calculation shows that the metric (\ref{ds6}), and the electromagnetic field tensor (\ref{F1}) and (\ref{gauge}) indeed satisfy all the field equations (\ref{eq1}) and (\ref{eq2}),  if  the metric function $H(r,x)$ satisfies the partial differential equation 
\be
r(\Lambda r^2+6)^2 {\frac {\partial ^{2}}{\partial {r}^{2}}}H
( r,x ) + 36 {\frac {\partial ^{2}}{
\partial {x}^{2}}}H ( r,x ) +(\Lambda r^2+18)(\Lambda r^2+6)\frac {\partial 
}{\partial r}H ( r,x)=0,\label{Feq}
\ee
as well as $r << \frac{1}{\sqrt{\Lambda}}$ . So, in what follows, we consider arbitrary small values for the cosmo-
logical constant $\Lambda$ to cover the whole range of the radial coordinate.
%where we consider $\nu$ to be the time coordinate in  (\ref{eq2}). 
%We also note that the non-diagonal $\mu=r,\,\nu=x$ component of equation (\ref{eq1}) is proportional to
%\be
%\frac{\partial}{\partial r}H(r,x)\frac{\partial}{\partial x}H(r,x) =0\label{rx}.
%\ee
To solve and find the solutions to equation (\ref{Feq}), we consider
\be
H(r,x)=1+ \alpha R(r)X(x),\label{sep}
\ee
where we determine the exact form of  the two functions $R(r)$ and $X(x)$ later, and $\alpha$ is a constant. Using equation (\ref{sep}) and substituting in equation (\ref{Feq}), we find that equation (\ref{Feq}) is completely separable into two ordinary differential equations for the functions $R(r)$ and $X(x)$. The two differential equations are given by
\be
(\frac{{r}^{5}}{36}{\Lambda}^{2}  +\frac{{r}^{3}}{3}\Lambda  +r    ){\frac {{\rm d}^{2}}{{\rm d}{r}^{2}}}R ( r
 )  +( \frac{{r}^{4}}{36}{\Lambda}^{2}     +\frac{2{r}^{2}\Lambda}{3}  +3     ) {\frac {\rm d}{
{\rm d}r}}R ( r )   
 +{p}^{2}rR \left( r \right) 
 =0\label{Req}
\ee
and
\be
{\frac {{\rm d}^{2}}{{\rm d}{x}^{2}}}X ( x
 ) -p^2 X(x)=0,\label{Xe}
\ee
respectively. We note that the constant $p$ in (\ref{Req}) and (\ref{Xe}), is the separation constant. We find the solutions to the differential equation (\ref{Req}) are given by
 \begin{eqnarray}
 R ( r )& =&{r_1}\, \left( 1+\frac{
\,\Lambda\,{r}^{2}}{6} \right) ^\xi
\left ({\mbox{$_2${F}$_1$}(\xi,\xi;2\xi -1;\,1+\frac{\,\Lambda\,{r}^{2}}{6})}+{\mbox{$_2${F}$_1$}^*(\xi,\xi;2\xi -1;\,1+\frac{\,\Lambda\,{r}^{2}}{6})}\right)\nonumber\\
&+&{r_2}\, \left( 1+\frac{\,\Lambda\,{r}^{2}}{6}
 \right) ^{2-\xi}
\left({\mbox{$_2$F$_1$}(2-\xi,2-\xi;3-2\xi;\,1+\frac{\,\Lambda\,{r}^{2}}{6})}+{\mbox{$_2$F$_1$}^*(2-\xi,2-\xi;3-2\xi;\,1+\frac{\,\Lambda\,{r}^{2}}{6})}\right),\nonumber\\
&&\label{Rsol}
 \end{eqnarray}
 where $r_1$ and $r_2$ are two arbitrary constants, $\mbox{$_2${F}$_1$}$ is the hypergeometric function, and 
 \be
\xi= \,{\frac {2\,\sqrt {\Lambda}-\sqrt {6\,{p}^{2}+4\,\Lambda}}{2\sqrt {\Lambda}}}\label{xi}.
 \ee
 In figure \ref{Fig1}, we plot the typical behaviour of the function $R(r)$ for several different values of the cosmological constant for $r_2=0$. As we notice the radial solutions are singular at $r=0$ and smoothly converge at large values of the radial coordinate. 

\begin{figure}[H]
\centering
\includegraphics[width=0.5\textwidth]{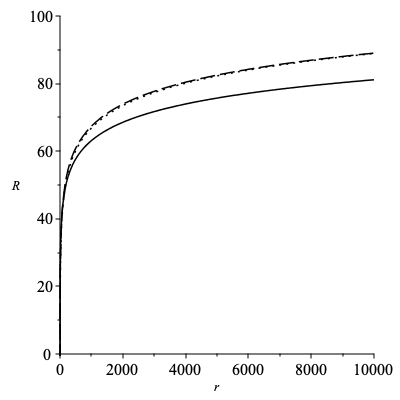}
\caption{The radial function $R(r)$ for different values of the cosmological constant $\Lambda=2$ (solid), $\Lambda=3$ (dash) and $\Lambda=4$ (dot), where we set $r_2=\frac{2}{\sqrt{6}}$, $r_1=0$, and $p=1$.}
\label{Fig2}
\end{figure}

Moreover, in figure \ref{Fig2}, we plot the typical behaviour of the function $R(r)$ for several different values of the cosmological constant for $r_1=0$. We notice from figure \ref{Fig2} that the radial solutions are convergent for $r\rightarrow \infty$ and divergent for $r\rightarrow 0$. However a detailed analysis of the solutions show that they are more singular at $r\rightarrow 0$ than solutions with $r_2=0$. However as we will discuss later, the solutions with  $r_2=0$ do not satisfy a necessary integral equation. The integral equation should be satisfied to correctly reduce the solutions to some exact solutions, where $r\rightarrow 0$  and  $\Lambda \rightarrow \infty$. So to avoid the inconsistency later, we set $r_1=0$, in equation (\ref{Rsol}).

The solutions to the differential equation (\ref{Xe}) are simply given by 
 \be
 X(x)=x_1 e^{-px}+x_2 e^{px},\label{Xs}
 \ee
 where $x_1$ and $x_2$ are two constants. In what follows, without loss of generality, we restrict the separation constant $p$ to be positive valued. So, to avoid a singular metric function for the large values of the positive coordinate $x$, we set $x_2=0$.
 \footnote{Alternatively choosing $x  \leq 0$ in the metric (\ref{ds6}), we should set  $x_1=0$ in (\ref{Xs}).} 

We note that one of the non-diagonal components of equation (\ref{eq1}) is zero, if we consider the small cosmological constant, such that  $\Lambda r^2$ is smaller than 1. Moreover all other components of equation (\ref{eq1}) are satisfied with the solutions (\ref{Rsol}) and (\ref{Xs}).

We can superimpose the different solutions with the different values of the separation constant $p$, to obtain the most general solutions to the differential equation (\ref{Feq}), as 
\begin{eqnarray}
 H(r,x)&=&1+\int _0^\infty dp P(p) e^{- px}\left( 1+\frac{
\,\Lambda\,{r}^{2}}{6} \right) ^{2-\xi}\nonumber\\
&\times&\left ({\mbox{$_2${F}$_1$}(2-\xi,2-\xi;3-2\xi ;\,1+\frac{\,\Lambda\,{r}^{2}}{6})}+{\mbox{$_2${F}$_1$}^*(2-\xi,2-\xi;3-2\xi ;\,1+\frac{\,\Lambda\,{r}^{2}}{6})}\right),\nonumber\\
&&\label{int}
 \end{eqnarray}
 where $P(p)$ is the weight function for each solution with a separation constant $p$. We take the double limits of the Fubini-Study metric (\ref{metr}), where $r\rightarrow 0$ and $\Lambda \rightarrow \infty$, while the combination $\Lambda r^2$ remain small quantity, which we denote by $6u$. The Fubini-Study metric reduces to
 \be
{\widehat {ds}}^2_{FS}=\frac{3du^2}{2\Lambda u(1+u)^2}+\frac{3u(d \psi +\cos(\theta)d \phi)^2}{2\Lambda(1+u)^2}+\frac{3u(d\theta^2+\sin(\theta)^2d\phi^2)}{2\Lambda(1+u)^2}.\label{redmetr}
 \ee 
 The very interesting point is that we find  the six-dimensional metric
 \be
 {\widehat {ds}}^2_6=-\frac{dt^{2}}{\widehat H(u,x)^{2}}+\widehat H(u,x)^{2/3}(dx^2+{\widehat {ds}}_{FS}^2),\label{metrasym}
 \ee
together with the Maxwell field tensor
\begin{eqnarray}
\widehat F_{tu}&=& - \frac{\gamma}{\widehat H(u,x)^2}\frac{\partial \widehat H(u,x)}{\partial u} ,   \\  \widehat F_{tx}&=&- \frac{\gamma}{\widehat H(u,x)^2}\frac{\partial \widehat H(u,x)}{\partial x} ,\label{Masym}
\end{eqnarray}
 are the exact solutions to the Einstein-Maxwell theory where 
 \be
 \widehat H(u,x)=1+\frac{\hat h }{(\frac{6u}{\Lambda}+x^2)^{3/2}},\label{Hasym}
 \ee
 where $\hat h$ is a constant. We now demand that the solutions (\ref{int}) for the metric function should reduce to the analytic result  (\ref{Hasym}), where we take the appropriate limits $r\rightarrow 0$ and $\Lambda \rightarrow \infty $. So, we find an integral equation for the weight function $P(p)$, which is given by
 \begin{eqnarray}
 \lim _{ r\rightarrow 0,\, \Lambda \rightarrow \infty}  &\int _0^\infty& dp P(p) e^{- px}\left( 1+\frac{
\,\Lambda\,{r}^{2}}{6} \right) ^{2-\xi}\nonumber\\
&\times&\left ({\mbox{$_2${F}$_1$}(2-\xi,2-\xi;3-2\xi ;\,1+\frac{\,\Lambda\,{r}^{2}}{6})}+{\mbox{$_2${F}$_1$}^*(2-\xi,2-\xi;3-2\xi ;\,1+\frac{\,\Lambda\,{r}^{2}}{6})}\right) \nonumber\\
&=& \frac{\hat h }{(\frac{6u}{\Lambda}+x^2)^{3/2}}.\label{inteq2}
 \end{eqnarray}
We should note that all occurrences of the $\Lambda r^2 = 6u$ are taken care of, in the double limits of $r \rightarrow 0,\,\Lambda \rightarrow \infty$. However $\Lambda$ also appears (without the multiplicative $r^2$) in the expressions (\ref{redmetr})-(\ref{inteq2}). So for these cases, we can’t perform the double limits, since $r^2$-terms are missing. So, we keep the track of $\Lambda$ in equations (\ref{redmetr})-(\ref{inteq2}).
 
\begin{figure}[H]
\centering
\includegraphics[width=0.5\textwidth]{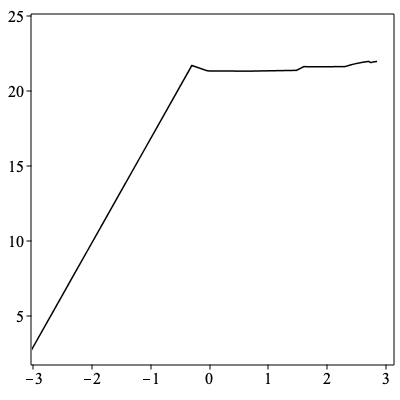}
\caption{The logarithmic plot of the metric function $H(r,x=0)$ versus $\log(r)$ where $\Lambda=3,\hat h=
6$.}
\label{Fig3}
\end{figure}

 To solve the very complicated integral equation (\ref{inteq2}), we notice that
 \be
\lim _{ r\rightarrow 0,\, \Lambda \rightarrow \infty} {\mbox{$_2${F}$_1$}(2-\xi,2-\xi;3-2\xi ;\,1+\frac{\,\Lambda\,{r}^{2}}{6})} = -\frac{2}{u} +
 {\cal O}(\ln u).\label{limitsF}
 \ee
 We find the solution to the integral equation (\ref{inteq2}) for $P(p)$,  is given by the constant function
 \be
 P(p)=- \frac{\hat h \Lambda}{36}.
 \ee 
 So, we get the most general solutions (\ref{int}) for the metric function $H(r,x)$, as
  \begin{eqnarray}
 H(r,x)&=&1- \frac{\hat h \Lambda}{36} \int _0^\infty dp  e^{- px}\left( 1+\frac{
\,\Lambda\,{r}^{2}}{6} \right) ^{2-\xi}\nonumber \\
&\times&
\left ({\mbox{$_2${F}$_1$}(2-\xi,2-\xi;3-2\xi ;\,1+\frac{\,\Lambda\,{r}^{2}}{6})}+{\mbox{$_2${F}$_1$}^*(2-\xi,2-\xi;3-2\xi ;\,1+\frac{\,\Lambda\,{r}^{2}}{6})}\right).\nonumber \\
&&\label{int2}
 \end{eqnarray}
 We should emphasis that if we considered $r_2=0$ solutions for the radial function in (\ref{Rsol}), then the hypergeometric function in the solutions in the limits of $r\rightarrow 0$ and $\Lambda \rightarrow \infty $ behaves as
 \be
\lim _{ r\rightarrow 0,\, \Lambda \rightarrow \infty} {\mbox{$_2${F}$_1$}(\xi,\xi;2\xi-1 ;\,1+\frac{\,\Lambda\,{r}^{2}}{6})} = 1.
 \ee
 So, there is not any $u$-dependence in the left hand side of the integral equation (\ref{inteq2}), if we chose the radial solutions with $r_2$=0 in (\ref{Rsol}). However the right side of the integral equation  (\ref{inteq2}) clearly depends on $u$.
 In figure \ref{Fig3}, we plot the result of numeric integration over $p$ for $H(r,x=0)$, where we set  $\Lambda=3,\hat h=6$. 
  
\begin{figure}[H]
\centering
\includegraphics[width=0.5\textwidth]{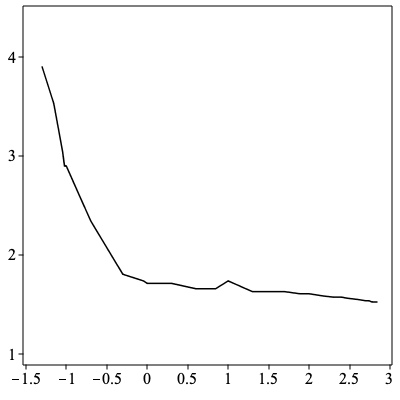}
\caption{The logarithmic plot of the metric function $H(r,x=1)$ versus $\log(r)$ where $\Lambda=3,\hat h=
6$.}
\label{Fig4}
\end{figure}

  Moreover, in figure  \ref{Fig4}, we plot the result of numeric integration over $p$ for $H(r,x=1)$, where we set  $\Lambda=3,\hat h=6$. We should note the sharp edges in figures \ref{Fig3} and \ref{Fig4}, are just the result of limitations of numerical calculation.  Increasing the coordinate $x$ leads to decreasing the overall value of the metric function exponentially.

 %%%%%%%%%%%%%%%%%%%%%%%%%%%%%%%%%%%%%%%%%%%%%%%%%%%%%%%%%%%%%%%%%%%%%%%%%%%%%%%%%%%%%%%%%%%%%%%%%%%%%%%%%%%%%%%%%%%%%%%%%%%%%%%%%%%%%%%%%%%%%%%%%%%%%%%%%%%%%%%%%%%%%%%%%%%%%%%%%%%%%%%%%%%%%%%%%%%%%%%%%%%%%
 \section{Embedding the four-dimensional Fubini-Study space in the seven dimensional Einstein-Maxwell theory}
 \label{sec:7D}
 We now try to embed the 4-dimensional Fubini-Study space (\ref{metr}) in the seven-dimensional spacetime, which is  given by
\begin{equation}
ds^{2}=-\frac{dt^{2}}{H(r,x)^{2}}+H(r,x)^{1/2}(dx^2+x^2 d\vartheta  ^2+ds_{FS}^2),
\label{ds7}
\end{equation}
where the coordinate $x\geq 0$ represents the radius of a two-dimensional disk with the angular coordinate $0 \leq \vartheta < 2\pi $. We consider the same components of the electromagnetic field tensor, as given by (\ref{gauge}).

The gravitational and Maxwell field equations are given by (\ref{eq1}) and (\ref{eq2}), respectively, where
$T_{\mu\nu}$ is the energy-momentum tensor for the electromagnetic field in 7-dimensions.
%\be
%T_{\mu\nu}=F_{\mu}^{\lambda}F_{\nu\lambda}-\frac{3}{4}F^2\label{T7}
%\ee

A straightforward but lengthy calculation shows that the only non-zero component of the Maxwell equations, is given by 
\be
r x (\Lambda r^2+6)^2 {\frac {\partial ^{2}}{\partial {r}^{2}}}H
( r,x ) + 36 r x  {\frac {\partial ^{2}}{
\partial {x}^{2}}}H ( r,x ) +x (\Lambda r^2+18)(\Lambda r^2+6)\frac {\partial 
}{\partial r}H ( r,x)+36\frac {\partial 
}{\partial x}H ( r,x)=0.\label{Feq7}
\ee
where we consider $\nu$ to be the time coordinate in (\ref{eq2}). 

We consider a separation of the variables in equation (\ref{Feq7}), as
\be
H(r,x)=1+ \alpha R(r)X(x),\label{sep7}
\ee
where we determine the exact form of  the two functions $R(r)$ and $X(x)$ later, and $\alpha$ is a constant. Using equation (\ref{sep7}) and substituting in equation (\ref{Feq7}), we find that equation (\ref{Feq7}) is completely separable into two ordinary differential equations for the functions $R(r)$ and $X(x)$. The two differential equations are given by
\be
({{r}^{5}}{\Lambda}^{2}  +12{{r}^{3}}\Lambda  +36 r   ){\frac {{\rm d}^{2}}{{\rm d}{r}^{2}}}R ( r
 )  +( {{r}^{4}}{\Lambda}^{2}     +24{{r^2}\Lambda} +108     ) {\frac {\rm d}{
{\rm d}r}}R ( r )   
 +{p}^{2}rR \left( r \right) 
 =0\label{Req7}
\ee
and
\be
36 x {\frac {{\rm d}^{2}}{{\rm d}{x}^{2}}}X ( x
 ) +36  {\frac {{\rm d}}{{\rm d}{x}}}X ( x
 )-p^2 x X(x)=0,\label{Xe7}
\ee
respectively,  where $p$  is the separation constant.

\begin{figure}[H]
\centering
\includegraphics[width=0.5\textwidth]{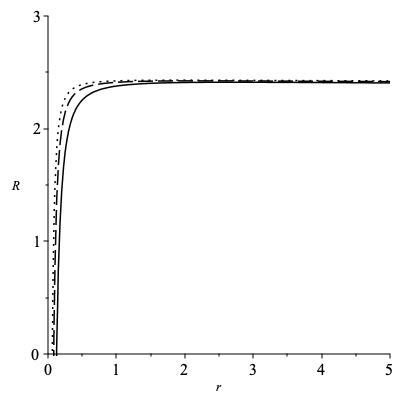}
\caption{The radial function $R(r)$ for different values of the cosmological constant $\Lambda=2$ (solid), $\Lambda=3$ (dash) and $\Lambda=4$ (dot), where we set $r_1=\frac{2}{\sqrt{6}}$, $s_2=0$, and $p=1$.}
\label{Fig5}
\end{figure}

\begin{figure}[H]
\centering
\includegraphics[width=0.5\textwidth]{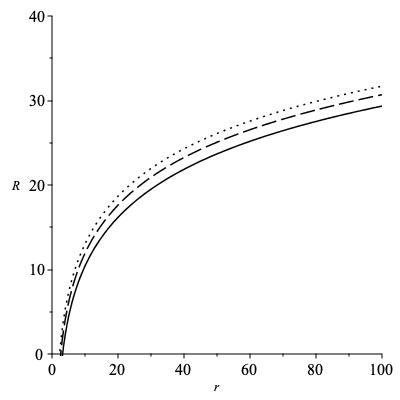}
\caption{The radial function $R(r)$ for different values of the cosmological constant $\Lambda=2$ (solid), $\Lambda=3$ (dash) and $\Lambda=4$ (dot), where we set $r_2=\frac{2}{\sqrt{6}}$, $s_1=0$, and $p=1$.}
\label{Fig6}
\end{figure}

We find the solutions to the differential equation (\ref{Req7}) are given by
 \begin{eqnarray}
 R ( r )& =&{s_1}\, \left( 1+\frac{
\,\Lambda\,{r}^{2}}{6} \right) ^\zeta
\left ({\mbox{$_2${F}$_1$}(\zeta,\zeta;2\zeta -1;\,1+\frac{\,\Lambda\,{r}^{2}}{6})}+{\mbox{$_2${F}$_1$}^*(\zeta,\zeta;2\zeta -1;\,1+\frac{\,\Lambda\,{r}^{2}}{6})}\right)\nonumber\\
&+&{s_2}\, \left( 1+\frac{\,\Lambda\,{r}^{2}}{6}
 \right) ^{2-\zeta}
\left({\mbox{$_2$F$_1$}(2-\zeta,2-\zeta;3-2\zeta;\,1+\frac{\,\Lambda\,{r}^{2}}{6})}+{\mbox{$_2$F$_1$}^*(2-\zeta,2-\zeta;3-2\zeta;\,1+\frac{\,\Lambda\,{r}^{2}}{6})}\right),\nonumber\\
&&\label{Rsol7}
 \end{eqnarray}
 where  $s_1$ and $s_2$ are two constants and 
 \be
\zeta= \,{\frac {12\,\sqrt {\Lambda}-\sqrt {6\,{p}^{2}+144\,\Lambda}}{12\sqrt {\Lambda}}}\label{xi7}.
 \ee
 In figure \ref{Fig5}, we plot the typical behaviour of the function $R(r)$ for several different values of the cosmological constant for $s_2=0$. We notice the radial solutions are singular at $r=0$ and smoothly converge at large values of the radial coordinate.
Moreover, in figure \ref{Fig6}, we plot the typical behaviour of the function $R(r)$ for several different values of the cosmological constant for $s_1=0$.

We notice from figure \ref{Fig6} that the radial solutions where $s_1=0$, are singular for $r=0$ too. A more detailed analysis of the radial solutions show that they are more singular than the  radial solutions with $s_2=0$. However as we noticed for the six dimensional solutions, the solutions with $s_2=0$ also can't satisfy a necessary integral equation. The integral equation guarantees that the most general solutions for the metric function approach correctly to some exact solutions of the theory, in the limits of $r\rightarrow 0$ and $\Lambda \rightarrow \infty$.
So, we set $s_1=0$, in equation (\ref{Rsol7}) to choose the proper radial solutions.

 The solutions to the differential equation (\ref{Xe7}) are given by 
 \be
 X(x)=y_1 I_0(\frac{px}{6})+y_2 K_0(\frac{px}{6}),\label{Xs7}
 \ee

where $y_1$ and $y_2$ are two constants, and $I_0$ and $K_0$ are the Bessel functions of the first and second kind, respectively. The Bessel function $I_0$ diverges where $x \rightarrow \infty$, while the Bessel function $K_0$ approaches to zero, where $x \rightarrow \infty$, where we restrict the separation constant $p$ to be a positive valued constant. To avoid a singular metric function for the large values of the positive coordinate $x$, we then set $y_1=0$ in (\ref{Xs7}).

We should note that the non-diagonal components of equation (\ref{eq1}) are zero. Moreover we find that all other components of equation (\ref{eq1}) are satisfied with the solutions (\ref{Rsol7}) and (\ref{Xs7}).

We superimpose the different solutions with the different values of the separation constant $p$, to obtain the most general solutions to the differential equation (\ref{Feq7}), as 
\begin{eqnarray}
 H(r,x)&=&1+\int _0^\infty dp Q(p) K_0(\frac{px}{6}) \left( 1+\frac{
\,\Lambda\,{r}^{2}}{6} \right) ^\zeta \nonumber\\
&\times&\left ({\mbox{$_2${F}$_1$}(2-\zeta,2-\zeta;3-2\zeta ;\,1+\frac{\,\Lambda\,{r}^{2}}{6})}+{\mbox{$_2${F}$_1$}^*(2-\zeta,2-\zeta;3-2\zeta ;\,1+\frac{\,\Lambda\,{r}^{2}}{6})}\right),\nonumber\\ \label{int7}
 \end{eqnarray}
 where $Q(p)$ is the weight function for each solution with a separation constant $p$ and $\zeta$ is given by (\ref{xi7}). As we did for six-dimensional solutions, to find the weight function $Q(p)$, we take the limit of the Fubini-Study metric (\ref{metr}), where $r\rightarrow 0$ and $\Lambda \rightarrow \infty$, while the combination $\Lambda r^2$ remain small quantity, which we denote by $6u$. The limit of the Fubini-Study metric is given by (\ref{redmetr}).
 
 Quite interestingly, we find  the seven-dimensional metric
 \be
 {\widetilde {ds}}^2_7=-\frac{dt^{2}}{\widetilde H(u,x)^{2}}+\widetilde H(u,x)^{1/2}(dx^2+x^2d\vartheta^2+{\widehat {ds}}_{FS}^2),\label{metrasym7}
 \ee
together with the Maxwell field tensor
\begin{eqnarray}
\widetilde F_{tu}&=& - \frac{\gamma}{\widetilde H(u,x)^2}\frac{\partial \widetilde H(u,x)}{\partial u} ,   \\  \widetilde F_{tx}&=&- \frac{\gamma}{\widetilde H(u,x)^2}\frac{\partial \widetilde H(u,x)}{\partial x} ,\label{Masym7}
\end{eqnarray}
 are the exact solutions to the Einstein-Maxwell theory where 
 \be
 \widetilde H(u,x)=1+\frac{\tilde h }{(\frac{6u}{\Lambda}+x^2)^{2}},\label{Hasym7}
 \ee
where $\tilde h$ is a constant. We now demand that the solutions (\ref{int7}) for the metric function should reduce to the analytic result  (\ref{Hasym7}), where we take the appropriate limits $r\rightarrow 0$ and $\Lambda \rightarrow \infty $. So, we find an integral equation for the weight function $Q(p)$, which is given by
 \begin{eqnarray}
 \lim _{ r\rightarrow 0,\, \Lambda \rightarrow \infty}  &\int _0^\infty& dp Q(p) K_0(\frac{px}{6}) \left( 1+\frac{
\,\Lambda\,{r}^{2}}{6} \right) ^\zeta\nonumber\\
&\times&\left ({\mbox{$_2${F}$_1$}(2-\zeta,2-\zeta;3-2\zeta ;\,1+\frac{\,\Lambda\,{r}^{2}}{6})}+{\mbox{$_2${F}$_1$}^*(2-\zeta,2-\zeta;3-2\zeta ;\,1+\frac{\,\Lambda\,{r}^{2}}{6})}\right) \nonumber\\
&=& \frac{\tilde h }{(\frac{6u}{\Lambda}+x^2)^{2}}.\label{inteq27}
 \end{eqnarray}
 To solve the  integral equation (\ref{inteq27}), we notice that
 
 \be
 \int _0 ^ N dp p^k K_0(\frac{px}{6})=\frac{3^{k+1}2^{2k}}{x^{1+k}}{\cal G}([[\frac{1-k}{2}, \frac{1-k}{2}],[1]],[[0],[]],\frac{144}{\chi}),\label{m7}
 \ee
 where ${\cal G}$ is the Meijer-G function \cite{GF}, and $\chi=N^2x^2$. We define the Meijer-G function and the Meijer-G equation in Appendix A. Comparing the integral equation (\ref{inteq27}) with (\ref{m7}) in the limit of $N \rightarrow \infty,  x\rightarrow 0$, we find that  $k=2$, and the weight function $Q(p)$ is
 \be
 Q(p)=- \frac{\tilde h \Lambda p^2}{2^{11} 3^4} .
 \ee 

\begin{figure}[H]
\centering
\includegraphics[width=0.5\textwidth]{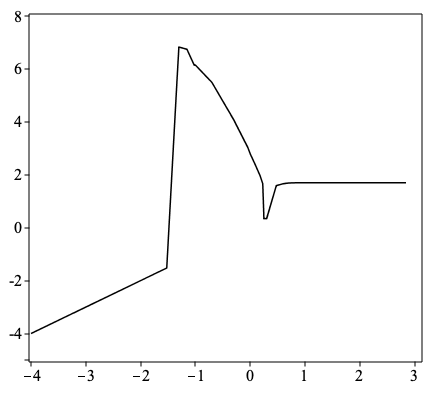}
\caption{The logarithmic plot of the metric function $H(r,x=1)$ versus $\log(r)$ where $\Lambda=2,\bar h_0=
1$.}
\label{Fig7}
\end{figure}

 So, we get the most general solutions (\ref{int7}) for the metric function $H(r,x)$, as
  \begin{eqnarray}
 H(r,x)&=&1- \bar h_0 \Lambda\int _0^\infty dp  p^2K_0(\frac{px}{6})\left( 1+\frac{
\,\Lambda\,{r}^{2}}{6} \right) ^\zeta\nonumber\\
&\times&
\left ({\mbox{$_2${F}$_1$}(2-\zeta,2-\zeta;3-2\zeta;\,1+\frac{\,\Lambda\,{r}^{2}}{6})}+{\mbox{$_2${F}$_1$}^*(\xi,\xi;2\xi -1;\,1+\frac{\,\Lambda\,{r}^{2}}{6})}\right),\nonumber\\ && \label{int27}
 \end{eqnarray}
 where $\zeta$ is given by (\ref{xi7} ) and $\bar h_0 = \frac{\tilde h }{2^{11} 3^4}$.
 In figure \ref{Fig7}, we plot the result of numerical integration over $p$ for $H(r,x=1)$, where we set  $\Lambda=3,\bar h_0=1$. 

\begin{figure}[H]
\centering
\includegraphics[width=0.5\textwidth]{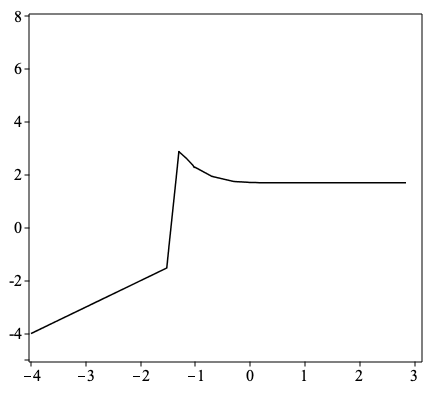}
\caption{The logarithmic plot of the metric function $H(r,x=10)$ versus $\log(r)$ where $\Lambda=3,\bar h_0=1
$.}
\label{Fig8}
\end{figure}

 Moreover, in figure  \ref{Fig8}, we plot the result of numeric integration over $p$ for $H(r,x=1)$, where we set  $\Lambda=3,\bar h_0=1$. 
We notice that increasing the coordinate $x$ leads to decreasing the overall value of the metric function in the middle region. We should note the sharp edges in figures \ref{Fig7} and \ref{Fig8}, are just the result of limitations of numerical calculation. 
 
\section{The Fubini-Study metric in higher-dimensional Einstein-Maxwell theory}
\label{sec:8D}
\subsection{Embedding the four-dimensional Fubini-Study space in the $D\geq 7$ dimensional Einstein-Maxwell theory}
In this section, inspired by the six dimensional solutions that we found in the previous section, we consider the $D\geq 7$-dimensional metric ansatz 
\be
ds_{D}^{2}=-H(r,x)^{-2}dt^{2}+H(r,x)^{2/(D-3)}(dx^2+x^2d\Omega_{D-6}+ds_{FS}^2),\label{mD}
\ee
where $d\Omega_{D-6}$ is the metric on unit sphere $S^{D-6}$ and $ds_{FS}^2$ is given by (\ref{metr}) . We consider the components of the Maxwell field tensor, as given by (\ref{gauge}).  We solve the field equations and find that all equations of motion are satisfied, pending on two differential equations for $R(r)$ and $X(x)$, where we separate $H(r,x)=1+hR(r)X(x)$. The differential equation for $R(r)$ in any dimensions $D\geq 7$ is given by (\ref{Req7}), while we find  the differential equation for $X(x)$ as 
\be
36x{\frac {d^{2}}{d{x}^{2}}}{\it X} \left( x \right) +36 \left( D-6
\right) {\frac {d}{dx}}{\it X} \left( x \right) - p^2 x {\it X} \left( x
\right) =0. \label{H2Deq}
\ee
We find the solutions to  (\ref{H2Deq}) are given by
\be
{\it X} \left( x \right) ={x_1}\,{x}^{\frac{7-D}{2}}
{{I_{\frac{D-7}{2}}}\left(\frac{px}{6}\right)}+{x_2}\,{x}^{\frac{7-D}{2}}
{{K_{\frac{D-7}{2}}}\left(\frac{px}{6}\right)}.
\ee
Hence,  we obtain the general superimposed solution in $D \geq 7$, as
\begin{eqnarray}
 H_D(r,x)&=&1+\int _0^\infty dp Q_D(p) {{K_{\frac{D-7}{2}}}\left(\frac{px}{6}\right)} \left( 1+\frac{
\,\Lambda\,{r}^{2}}{6} \right) ^\zeta \nonumber\\
&\times&\left ({\mbox{$_2${F}$_1$}(2-\zeta,2-\zeta;3-2\zeta ;\,1+\frac{\,\Lambda\,{r}^{2}}{6})}+{\mbox{$_2${F}$_1$}^*(2-\zeta,2-\zeta;3-2\zeta ;\,1+\frac{\,\Lambda\,{r}^{2}}{6})}\right),\nonumber\\ \label{intD}
 \end{eqnarray}
where $Q_{D}(p)$ is the weight function.
To find the weight function $Q_{D}(p)$ , we consider the limit of the Fubini-Study metric (\ref{metr}), where  $r \rightarrow 0$ and $\Lambda \rightarrow \infty$, while the combination $6u=\Lambda r^2$ remains finite and small. 
We find that the $D$-dimensional metric
 \be
 {\widetilde {ds}}^2_D=-\frac{dt^{2}}{\widetilde H_D(u,x)^{2}}+\widetilde H_D(u,x)^{\frac{2}{D-3}}(dx^2+x^2d\Omega_{D-6}^2+{\widehat {ds}}_{FS}^2),\label{metrasymD}
 \ee
together with the Maxwell field tensor components (\ref{Masym7}),  are the exact solutions to the $D$-dimensional Einstein-Maxwell theory where 
 \be
 \widetilde H_D(u,x)=1+\frac{\tilde h_D }{(\frac{6u}{\Lambda}+x^2)^{\frac{D-3}{2}}},\label{HasymD}
 \ee
 where ${\tilde h_D }$ is a constant. 
 Similar to previous cases in six and seven dimensions, we now demand that the solutions (\ref{intD}) for the metric function should reduce to the analytic result  (\ref{HasymD}), where we take the appropriate limits $r\rightarrow 0$ and $\Lambda \rightarrow \infty $. We find an integral equation for the weight function $Q_D(p)$, which is given by
 \begin{eqnarray}
 \lim _{ r\rightarrow 0,\, \Lambda \rightarrow \infty}  &\int _0^\infty& dp Q_D(p) K_{\frac{D-7}{2}}(\frac{px}{6}) \left( 1+\frac{
\,\Lambda\,{r}^{2}}{6} \right) ^\zeta\nonumber\\
&\times&\left ({\mbox{$_2${F}$_1$}(2-\zeta,2-\zeta;3-2\zeta ;\,1+\frac{\,\Lambda\,{r}^{2}}{6})}+{\mbox{$_2${F}$_1$}^*(2-\zeta,2-\zeta;3-2\zeta ;\,1+\frac{\,\Lambda\,{r}^{2}}{6})}\right) \nonumber\\
&=& \frac{\tilde h_D }{(\frac{6u}{\Lambda}+x^2)^{\frac{D-3}{2}}}.\label{inteq2D}
 \end{eqnarray}
 To solve the  integral equation (\ref{inteq2D}), we notice that for $D=7+2n,\, n=0,1,2,\cdots$
 \be
 \int _0 ^ N dp p^{2+2n} K_n(\frac{px}{6})=-\frac{2^{4n+6}3^{2n+3}}{x^{3+2n}}{\cal G}([[-\frac{3n+1}{2},- \frac{n+1}{2}],[1]],[[0],[]],\frac{144}{\chi}),\label{mD1}
 \ee
 where ${\cal G}$ is the Meijer-G function \cite{GF},
 and $\chi=N^2x^2$. Comparing the integral equation (\ref{inteq2D}) with (\ref{mD1}) in the limit of $N \rightarrow \infty,  x\rightarrow 0$, we find that the weight function $Q_D(p)$ is given by
 \be
 Q_D(p)=- \frac{\tilde h_D \Lambda p^{2n+2}}{(n+2)2^{4n+10}3^{2n+4}} .
 \ee 
 To summarize, we find the metric function $H_D(r,x)$ for $D=7+2n$ is given by
\begin{eqnarray}
 H_{7+2n}(r,x)&=&1-\bar h_n \Lambda \int _0^\infty dp p^{2n+2} {{K_{{n}{}}}\left(\frac{px}{6}\right)} \left( 1+\frac{
\,\Lambda\,{r}^{2}}{6} \right) ^\zeta \nonumber\\
&\times&\left ({\mbox{$_2${F}$_1$}(2-\zeta,2-\zeta;3-2\zeta ;\,1+\frac{\,\Lambda\,{r}^{2}}{6})}+{\mbox{$_2${F}$_1$}^*(2-\zeta,2-\zeta;3-2\zeta ;\,1+\frac{\,\Lambda\,{r}^{2}}{6})}\right),\nonumber\\ \label{intDD}
 \end{eqnarray}
where $\bar h_n=\frac{\tilde h_D}{(n+2)2^{4n+10}3^{2n+4}}$.

\begin{figure}[H]
\centering
\includegraphics[width=0.5\textwidth]{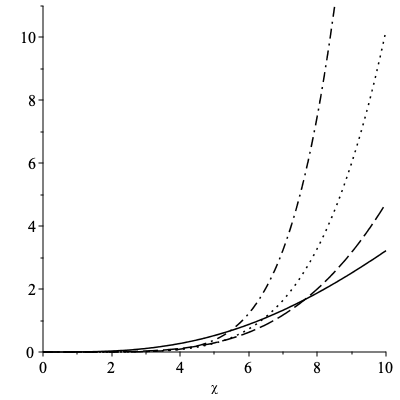}
\caption{The function ${\cal F}_n(\chi)$ versus $\chi$, for $n=0$ (solid), $n=1$ (dash), $n=2$ (dot) and $n=3$ (dashdot).}
\label{Fig9}
\end{figure}

On the other hand, for $D=8+2n,\,n=0,1,2,\cdots $, we find
  \be
 \int _0 ^ N dp p^{3+2n} K_{n+\frac{1}{2}}(\frac{px}{6})=\frac{2^{4n+8}3^{2n+4}}{x^{4+2n}(2n+3)(2n+7)}{\cal F}_n({\chi}),\label{mDD}
 \ee
where $\chi=Nx$ and the function ${\cal F}_n(\chi)$ is given by
\begin{eqnarray}
{\cal F}_n(\chi)&=&\,{2}^{-4-2\,n}{3}^{-\frac{7}{2}-n}{\chi}^{\frac{7}{2}+n}
{\mbox{$_1$F$_2$}(\frac{7}{4}+\frac{n}{2};\,\frac{1}{2}-n,\frac{11}{4}+\frac{n}{2};\,{\frac {{\chi}^{2}}{144}})}
\Gamma \left( n+\frac{1}{2} \right) n\nonumber\\
&+&\,{2}^{-6-6\,n}{3}^{-\frac{11}{2}-3\,n}{\chi}^{
\frac{9}{2}+3\,n}
{\mbox{$_1$F$_2$}(\frac{9}{4}+\frac{3}{2}\,n;\,\frac{3}{2}+n,{\frac{13}{4}}+\frac{3}{2}\,n;\,{\frac {{\chi}^{2}}{144}})}
\Gamma \left( -n-\frac{1}{2} \right) n\nonumber\\
&+&\,{2}^{-5-2\,n}{3}^{-\frac{5}{2}-n}{\chi}^{\frac{7}{2}
+n}
{\mbox{$_1$F$_2$}(\frac{7}{4}+\frac{n}{2};\,\frac{1}{2}-n,\frac{11}{4}+\frac{n}{2};\,{\frac {{\chi}^{2}}{144}})}
\Gamma \left( n+\frac{1}{2} \right) \nonumber\\
&+&\,{2}^{-7-6\,n}{3}^{-\frac{11}{2}-3\,n}7{\chi}^{\frac{9}{2}
+3\,n}
{\mbox{$_1$F$_2$}(\frac{9}{4}+\frac{3}{2}\,n;\,\frac{3}{2}+n,{\frac{13}{4}}+\frac{3}{2}\,n;\,{\frac {{\chi}^{2}}{144}})}
\Gamma \left( -n-\frac{1}{2} \right).\nonumber\\
&&
\end{eqnarray}
We plot the typical behaviour of ${\cal F}_n(\chi)$ for several values of $n$ in figure \ref{Fig9}. 
In the limit of $x\rightarrow 0$, we always can choose an $N\rightarrow \infty$, such that ${\cal F}_n(\chi)=1$.  Comparing the integral equation (\ref{inteq2D}) with (\ref{mDD}) in the limit of $N\rightarrow \infty$, $x\rightarrow 0$, we find that the weight function $Q_D(p)$ is 
\be
 Q_D(p)=- \frac{\tilde h_D \Lambda (2n+3)(2n+7) p^{2n+3}}{(2n+5)2^{4n+10}3^{2n+5}} .
\ee
To summarize, we find the metric function $H_D(r,x)$ for $D=8+2n$ is given by
\begin{eqnarray}
 H_{8+2n}(r,x)&=&1-\bar h_n \Lambda \int _0^\infty dp p^{2n+3} {{K_{{n+\frac{1}{2}}{}}}\left(\frac{px}{6}\right)} \left( 1+\frac{
\,\Lambda\,{r}^{2}}{6} \right) ^\zeta \nonumber\\
&\times&\left ({\mbox{$_2${F}$_1$}(2-\zeta,2-\zeta;3-2\zeta ;\,1+\frac{\,\Lambda\,{r}^{2}}{6})}+{\mbox{$_2${F}$_1$}^*(2-\zeta,2-\zeta;3-2\zeta ;\,1+\frac{\,\Lambda\,{r}^{2}}{6})}\right),\nonumber\\ \label{intDD8}
 \end{eqnarray}
where $\bar h_n=\frac{\tilde h_D (2n+3)(2n+7)}{(2n+5)2^{4n+10}3^{2n+5}}$.

\subsection{Any cosmological solutions in $D\geq 6$ dimensional Einstein-Maxwell theory?}

We note that for all the solutions presented so far, there is no cosmological constant in $D\geq 6$ Einstein-Maxwell theory. Of course, it is an obvious objective to try and find the cosmological solutions for $D\geq 6$ Einstein-Maxwell theory. In fact, in references \cite{kumar, me2}, the authors found that adding a linear term of the time coordinate $t$ to the convoluted solutions makes them cosmological solutions to the Einstein-Maxwell theory in presence of the cosmological constant. Inspired by these works, in this section, we show that there are no non-trivial solutions to the $D\geq 6$ dimensional Einstein-Maxwell theory with a cosmological constant  ${\bold \Lambda} $, if we consider a special anasatz for the metric, which is given by (\ref{mDLambda})  with a time-dependent metric function. Of course, we don't roll out finding other possible cosmological solutions with more sophisticated ansatzes. We note that ${\bold \Lambda} $ is different from $\Lambda$ that appear in the Fubini-Study metric (\ref{metr}). We consider the following ansatz for the $D\geq 6$ dimensional metric 
\be
ds_{D}^{2}=-H(t,r,x)^{-2}dt^{2}+H(t,r,x)^{2/(D-3)}(dx^2+x^2d\Omega_{D-6}+ds_{FS}^2),\label{mDLambda}
\ee

where $d\Omega_{D-6}$ is the metric on unit sphere $S^{D-6}$ and $ds_{FS}^2$ is given by (\ref{metr}) . We consider the components of the Maxwell field tensor, as given by (\ref{gauge}). We note that the metric function $H(t,r,x)$ depends on $t$, as well as $r$, and $x$.  The Einstein field equations are given by
\be
G_{\mu\nu}+{\bf \Lambda}g_{\mu\nu}=T_{\mu\nu}
\ee
together with the Maxwell field equations  (\ref{eq2}). The Maxwell field equations (\ref{eq2}) imply that $H(t,r,x)$ should be separated as
\be
H(t,r,x)={\bold T}(t)+{\bold R}(r){\bold X}(x),\label{sep3}
\ee
where ${\bold T}$, ${\bold R}$ and ${\bold X}$ are three functions of $t$, $r$ and $x$, respectively. After separation of coordinates, 
the Maxwell equations lead to
\begin{eqnarray}
{\bold R}(r)&=&R(r) ,\\
{\bold X}(x)&=&X(x),
\end{eqnarray}
where $R(r)$ and $X(x)$ are the solutions found in the previous sections in different dimensions. Substituiting $H(t,r,x)$ as given by equation (\ref{sep3}) in the Einstein field equations, leads to
\begin{eqnarray}
{\bold T}(t)&=&(D-3)\sqrt{\frac{2{\bold \Lambda}}{(D-1)(D-2)}}t+t_0,\\
\Lambda&=&0.\label{cont}
\end{eqnarray}
Obviously, equation (\ref{cont}) is inconsistent with the non-trivial Fubini-Study metric (\ref{metr}), where $\Lambda \neq 0$.

\section{Concluding Remarks}
\label{sec:con}

We construct  solutions to the Einstein-Maxwell theory in six and higher dimensions. The solutions are based on uplifting the four dimensional Kahler Fubini-Study manifold into higher dimensional gravity coupled to the electromagnetic field.  To consistently construct the solutions, we superimpose the different solutions into an integral with an unknown weight function. We find exactly the form of the weight function, with matching the solutions to some exact solutions, in the proper limit of some parameters and coordinates.  The numerical results for the metric function show that the solutions are regular everywhere, except on a single point $r=0$. We expect the singularity may be located inside a higher dimensional regular hyper-surface, if  we consider  more sophisticated dependence of  the metric function on other coordinates.
We also show explicitly that including the cosmological constant in the theory implies trivialization of the Fubini-Study metric. Hence we can't uplift the Fubini-Study metric (with already an intrinsic cosmological constant) into a cosmological solution in higher dimensional Einstein-Maxwell theory. We should stress that the absence of the cosmological solutions is based only on considering a special anasatz for the metric, which is given by (\ref{mDLambda}) with a time-dependent metric function. Of course, we don’t roll out finding other possible cosmological solutions with more sophisticated ansatzes.

\vspace*{0.8cm}

\bigskip
{\large \bf Acknowledgments}
\vspace*{0.5cm}

This work was supported by the Natural Sciences and Engineering Research
Council of Canada. All data generated or analysed during this study are included in this published article.

\vspace*{0.4cm}

\bigskip
{\large \bf Appendix A: The Meijer-G function}\pagebreak
\vspace*{0.5cm}

The Meijer G-function G is defined in terms of $\Gamma$ functions by
\begin{eqnarray}
&&{\cal G}([[a_1,\cdots ,a_m],[b_1,\cdots ,b_n]],[[c_1,\cdots ,c_p],[d_1,\cdots ,d_q]],z] = \nonumber\\
&=&\frac{1}{2\pi i}\oint _C \frac{\prod_{i=1}^m \Gamma (1-a_i+y)}{\prod_{j=1}^n \Gamma (b_j-y)}
\frac{\prod_{k=1}^p \Gamma (c_k-y)}{\prod_{l=1}^q \Gamma (1-d_l+y)}z^w dw,\label{GF}
\end{eqnarray}
where $m + n \leq p + q$ and $C$ is the integration contour on the complex $z$-plane. The Meijer
G-function (\ref{GF})  is the solution to the differential equation
\begin{equation}
\{(-1)^{n-p-m}z\prod_{i=1}^m(z\frac{d}{dz}-a_i+1)
\prod_{j=1}^n(z\frac{d}{dz}-b_j+1)-
\prod_{k=1}^p(z\frac{d}{dz}-c_k)
\prod_{l=1}^q(z\frac{d}{dz}-d_l)\}{\cal G}=0.
\end{equation}

\end{document}